\documentclass[12pt,preprint]{aastex}

\shortauthors{Feuchtgruber et al.}
\shorttitle{New wavelength determinations}
\slugcomment{Submitted March 2001}
\begin{document}

\title{New rest wavelength determinations for 7 mid-infrared fine structure 
       lines by ISO-SWS\footnote
        {Based on observations made with ISO, an project of ESA with
         participation of ISAS and NASA, and the SWS, a joint project of SRON
         and MPE with contributions from KU Leuven, Steward Observatory and
         Phillips Laboratory.}}

\author{H. Feuchtgruber, D. Lutz}
\affil{Max-Planck-Institut f\"ur extraterrestrische Physik,
                 Postfach 1312, D--85741 Garching, Germany}

\email{fgb@mpe.mpg.de}

\author{D. A. Beintema} 
\affil{SRON Laboratory for Space Research, P.O. Box 800, 
                 9700 AV Groningen, The Netherlands}

\begin{abstract}
 
Observations of the planetary nebulae NGC\,6302, NGC\,6543 and NGC\,7027 
by the Short Wavelength Spectrometer (SWS) on
board the Infrared Space Observatory (ISO) have been used to determine 
rest wavelengths of spectral lines. We report 
on improved accuracies for wavelengths of 7 mid-infrared ionic fine structure 
lines.

\end{abstract}


\section{Introduction}

Mid infrared atomic fine structure lines are important probes for a variety of 
astrophysical environments like planetary nebulae (e.g., Beintema 
et al. 1996), shocked regions (e.g., Oliva et al. 1999) and 
galaxies (e.g., Lutz et al. 2000). Due to their low magnetic dipole transition 
probabilities the lines are usually optically thin ($\tau = 10^{-2}$ to a few)
and measurements of these lines give handles on luminosities and
spatial distribution of abundant atoms and ions. In addition they provide 
also valuable information on the density, temperature and ionization 
structure of targets 
with little or no dependence on dust extinction that hampers observations in 
the visible. Accurate laboratory measurements of highly ionized species are 
difficult and line rest wavelengths are mostly based on energy level 
differences reconstructed from UV and optical spectroscopy. The knowledge of
accurate rest wavelengths of these lines will enhance their diagnostic value
for many spectroscopic studies in the mid-infrared wavelength range.

New wavelength determinations by the Short Wavelength 
Spectrometer (SWS; de Graauw et al. 1996) on board the Infrared 
Space Observatory (ISO; Kessler et al. 1996) have been reported 
by Feuchtgruber et al. 1997 (hereafter paper I). 
The observations therein were taken early during the ISO mission 
and the subsequent analysis was focused mainly on bright lines. Long 
integration observations on NGC\,6302, NGC\,6543 and NGC\,7027 obtained later 
during the operational lifetime of ISO revealed another 7 lines where SWS 
data provide improved accuracies on their rest wavelengths. Thus, the 
present work represents an extension of paper I towards fainter lines. 
Transitions from [Cl IV], [Cl V], [Ca V], Ca [VII] 
and [Al VI] have been measured in SWS AOT06 grating mode, while the 
21.8 $\mu$m line of [Ar III] has been detected with the SWS Fabry-Perot (FP), 
improving significantly on its accuracy as compared to the SWS grating 
observation reported in paper I. 

The accuracy of the SWS wavelength calibration has been established initially 
by Valentijn et al. 1996 down to an uncertainty of 0.5-1.0 steps
of the grating scanner. This value, translated from its instrumental unit into
wavelength, typically corresponds to $\lambda/\Delta\lambda=1\times 10^{4}$.
The nominal 2 steps uncertainty of the FP at 21.8 micron correspond 
to $\lambda/\Delta\lambda=1\times 10^{5}$. While the FP wavelength calibration
has proven to be stable throughout the whole ISO mission, the SWS grating
wavelength calibration has shown a slight shift in time ($<0.1$ step/month). 
Regular calibration observations of the internal SWS grating calibration 
source and external astronomical wavelength calibration targets have however
allowed to maintain the initial accuracy by interpolation between the 
various measurements. For more details regarding the accuracy of
wavelengths in SWS observations see paper I and 
Valentijn et al. 1996.

\begin{deluxetable}{ccccccrc}
   \tablewidth{0pc}
   \tabletypesize{\scriptsize}
   \tablecaption{Summary of new wavelength determinations \label{newlambda}}
\tablehead{
\colhead{Ion} & \colhead{Transition} & \colhead {Vac. Wavelength$^*$} & 
\colhead{Rest wavenumber$^*$} &\colhead{$\chi_{lower}$} & 
\colhead{$\chi_{upper}$} &\colhead{Obs. Date} &\colhead{Target}\\
\colhead{}&\colhead{$u-l$}&\colhead{[$\mu$m]}&\colhead{[cm$^{-1}$]} &
\colhead{[eV]}&\colhead{[eV]}&\colhead{}&\colhead{}} 

\startdata
    [Al VI] & $^3$P$_1$-$^3$P$_2$ & 3.65971$\pm$(62) &
    2732.46$\pm$(46) & 153.83& 190.48& 1997-Feb-20$^a$ & NGC\,6302 \\

    [Ca VII] & $^3$P$_2$-$^3$P$_1$ & 4.0858$\pm$(12) &
    2447.5$\pm$(7) & 108.78 & 127.20& 1997-Feb-20$^a$ & NGC\,6302 \\

    [Ca V] & $^3$P$_1$-$^3$P$_2$  & 4.15937$\pm$(62) &
    2404.21$\pm$(36) & 67.27 & 84.50& 1996-Oct-19$^a$ & NGC\,7027 \\

    [Cl V] & $^2$P$_{3/2}$-$^2$P$_{1/2}$  & 6.70667$\pm$(62) &
    1491.05$\pm$(14) & 53.46 & 67.82 & 1996-Oct-19$^a$ & NGC\,7027 \\

    [Cl IV] & $^3$P$_{2}$-$^3$P$_{1}$  & 11.7619$\pm$(11) &
    850.203$\pm$(79) & 39.61 & 53.46 & 1996-Oct-19$^a$ & NGC\,7027 \\

    [Cl IV] & $^3$P$_{1}$-$^3$P$_{0}$  & 20.3107$\pm$(21) &
    492.351$\pm$(51) & 39.61 & 53.46 & 1997-Feb-20$^a$ & NGC\,6302 \\

    [Ar III] & $^3$P$_0$-$^3$P$_1$  & 21.8302$\pm$(3) &
    458.081$\pm$(6) & 27.63 & 40.74 & 1997-Feb-11$^b$ & NGC\,6543 \\

\tablenotetext{a}{Grating scan}
\tablenotetext{b}{Fabry-Perot line scan} 
\tablenotetext{*}{Numbers in brackets give errors on last decimals} 
\tablecomments{$\chi_{lower}$ and $\chi_{upper}$ denote excitation and 
                 ionization potentials}
\enddata
\end{deluxetable}

\section{Observations and data reduction}

We have analyzed observations, taken in SWS AOT06 grating mode of the 
planetary nebulae NGC\,6302 and NGC\,7027. NGC\,6543 has been observed in 
SWS AOT07 Fabry-Perot mode. The rich SWS spectra of these targets are 
discussed e.g. by Beintema et al. 1996, 
Pottasch et al. 1996, Beintema \& Pottasch 1999 and
Pottasch \& Beintema 1999. Here, we focus exclusively
on wavelength determination and report only about those lines 
observed by SWS for which we could improve on the accuracy of the rest 
wavelengths. 

The data reduction of all measurements was done using the SWS Interactive
Analysis software (SIA; Wieprecht et al. 1998) package on 
ISO pipeline products of version OLP 8.4. 
Particular attention has been payed to remove instrumental fringing in the 
wavelength range between 12 $\mu$m and 29 $\mu$m by the dedicated fourier 
filtering modules of SIA.
The same procedure was used as for paper I. 
Gaussian fits to the spectral lines provided the 
center wavelengths,
which were corrected for the heliocentric radial velocity and for the mean 
ISO radial velocity. Typical velocity changes during an observation were
less than 0.3 km/s, so could be neglected.
The observed profiles of all lines with new wavelengths are presented in 
Fig.~1. 

\subsection{Use of reference lines}

Small pointing errors and spatial inhomogeneities of the targets can induce 
apparent velocity shifts in SWS grating data.
We have corrected for this effect using reference lines of accurately known 
wavelengths, observed in the same SWS aperture during the same observation.
Details of this strategy are given in paper I and will not be repeated here.
Although at similar accuracy, some reference lines with rest 
wavelengths determined in paper I have been used (see Table~\ref{reflambda}), 
to improve on fixing the reference frame and to illustrate the good agreement 
with the more accurate reference lines. 

The new wavelength determinations together with some other line properties are
summarized in Table~\ref{newlambda}. The framework of fine structure 
and hydrogen or helium recombination reference
lines which were used for the indvidual observations of our targets is listed
in Table~\ref{reflambda}. The offsets 
reflect the pointing error and the effects of the spatial structure of the 
source of the particular observation. 
Together with the uncertainty of the literature wavenumber, the 
`SWS uncertainty' in Table~\ref{reflambda} is propagated into the uncertainty 
of new wavelength determinations that are tied to this line.

Apart from literature wavenumbers, Table~\ref{reflambda} also includes a
rest wavenumber for NGC\,6302 which was determined during the independent 
observation by the SWS FP on NGC\,6543.

\begin{figure}
\plotone{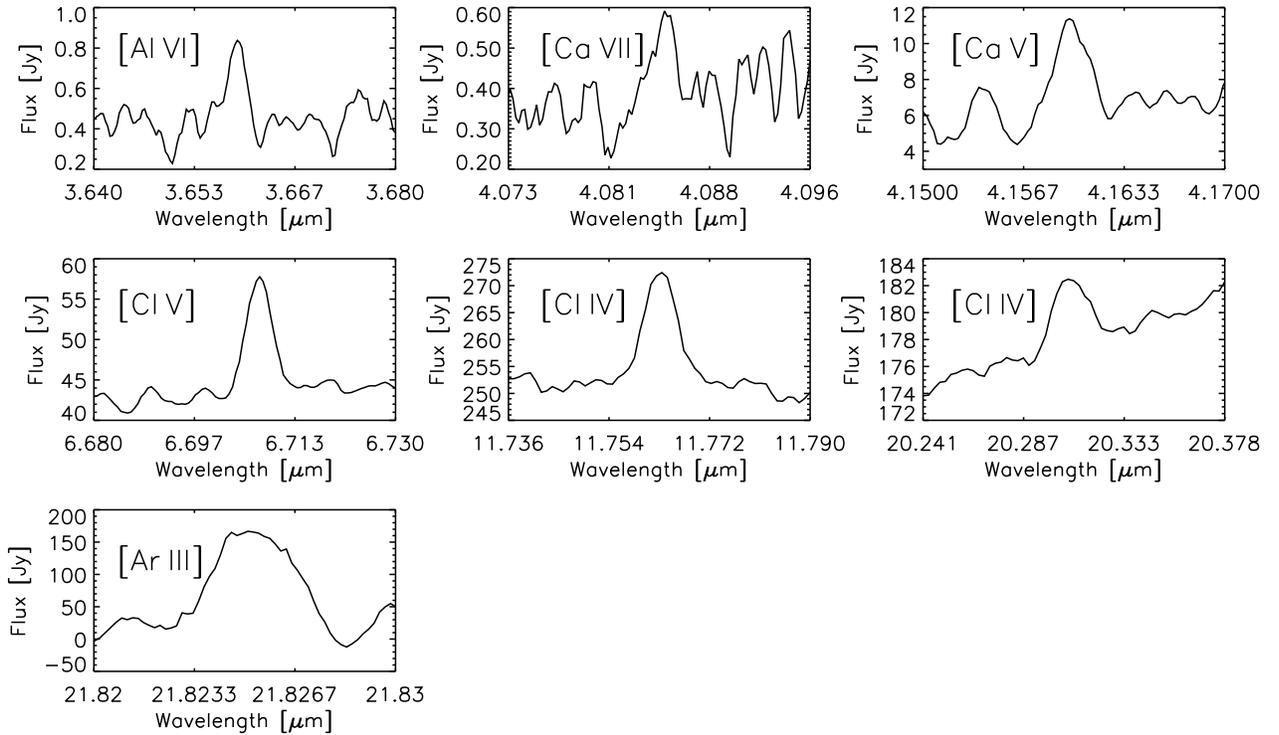}
\caption{Observed fine structure lines: [Al VI], [Ca VII] and [Cl IV]
        (20.3 $\mu$m) from
         NGC\,6302; [Ca V], [Cl V] and [ClIV] (11.76 $\mu$m) from NGC\,7027;
         [Ar III] from NGC\,6543}
\end{figure}

\subsection{NGC\,6302}

The SWS grating observation of NGC\,6302 has been carried out on 
1997 February 20. The total integration time has been 8532 s. Beintema \& Pottasch (1999) 
have presented the first analysis of this AOT SWS06 observation and provided 
an extensive
line list for the SWS wavelength range. 
From this observation we derived improved rest wavelengths for the three 
lines reported in table~\ref{newlambda}. The [CL IV] line has escaped the 
analysis by Beintema \& Pottasch, who quote an upper limit to the line flux. 
For the faint lines of [Al VI] and [Ca VII] the uncertainties of the 
wavelength determinations are dominated by detector noise.

From the analysis of the set of reference lines 
(see table~\ref{reflambda}), measured at 
similar wavelength and with the same SWS aperture, the observed lines 
in NGC\,6302 
had to be shifted by $-4 km s^{-1}$ for [Cl IV] and $-26 km s^{-1}$
for [Al VI] and [Ca VII] with respect to the heliocentric velocity 
of $-39\pm2 km s^{-1}$ given by Acker et al. (1992), probably due 
to a pointing error. Note that a pointing error may translate into 
different shifts for the short wavelength and long wavelength sections of SWS 
because their beam profiles are centered on slightly different positions on 
the sky (Salama 2000), and since the velocity shift corresponding to a given 
angular shift differs for different order and aperture combinations.   

\subsection{NGC\,6543}

The ionic line from [Ar\,III] at 21.83 $\mu$m has been detected by the ISO SWS 
FP during an observation of 6750 s on 1997 February 11. 
The wavelengths from the [Ne\,III] line at 
15.5 $\mu$m and of the [S\,III] line at 18.71 $\mu$m have been remeasured 
during the same observation. Both reproduce the results from paper I within 
their quoted uncertainties, confirming the assumption of sufficient velocity
and spatial averaging across the target through the SWS slit sizes, since the 
[Ne\,III] wavelength has been originally determined on NGC\,7027 and the 
[S\,III] line has been observed at a different position angle of the SWS slit
on NGC\,6543.
The heliocentric velocity correction of -66.1$\pm$0.4 km/s for NGC\,6543 has 
been taken from Schneider et al. (1983).
The determined rest wavelength for the [Ar\,III] transition is in agreement 
with the SWS grating result from paper I, however its accuracy is now 
significantly improved. 

\subsection{NGC\,7027}

The planetary nebula NGC\,7027 has been observed with SWS on 1996 October 19.
The total integration time of the AOT06 observation has been 7962 s. 
New wavelength have been 
determined for ionic lines of [Ca V], [Cl IV] and [Cl V]. The data are 
presented in Fig.~1 and their new rest wavelengths in Table~\ref{newlambda}.
For this observation we found a mean shift of $+24 km s^{-1}$ of 
the reference lines (Table~\ref{reflambda}) with respect to the 
heliocentric velocity of $+8.8\pm0.6 km s^{-1}$ from Schneider 
et al. (1983).

\begin{deluxetable}{cccccccc}
   \tablewidth{0pc}
\tabletypesize{\scriptsize}
   \tablecaption{Summary of reference wavenumbers. See section 2.1 and paper I
                 for explanations.    \label{reflambda}}
\tablehead{
\colhead{Obs. Date} &\colhead{Target} & \colhead{Ion} & \colhead{Transition} & 
\colhead{Ref. Wavenumber} &\colhead{Offset$^a$} & 
\colhead{SWS uncertainty$^b$} &\colhead{Reference}\\
\colhead{}&\colhead{}&\colhead{}&\colhead{$u-l$} &
\colhead{[cm$^{-1}$]}&\colhead{[cm$^{-1}$]}&\colhead{[cm$^{-1}$]}&\colhead{}} 

\startdata
    1997-Feb-20 & NGC\,6302 &    H$_2 \,O(5) $   & 1 - 0  & 3091.20$\pm \le$0.03& 0.05 & 0.19 &
 Black \& Dishoek 1987\\
    1997-Feb-20 & NGC\,6302 &    HI  & 9 - 5 & 3033.07$\pm \le$0.03& 0.08 & 0.19 &
 van Hoof \& Verner 1997\\
    1997-Feb-20 & NGC\,6302 &    HI  & 8 - 5 & 2673.40$\pm \le$0.03& 0.14 & 0.28 &
 van Hoof \& Verner 1997\\
    1997-Feb-20 & NGC\,6302 &    [Mg IV] & $^2$P$_{1/2}$-$^2$P$_{3/2}$ & 2228.82$\pm$0.15& -0.02 & 0.15 &
 Feuchtgruber et al. 1997\\
    1997-Feb-20 & NGC\,6302 &    [Ar VI] & $^2$P$_{3/2}$-$^2$P$_{1/2}$ & 2207.74$\pm$0.15& 0.02 & 0.15 &
 Feuchtgruber et al. 1997\\
    1997-Feb-20 & NGC\,6302 &    [K III] & $^2$P$_{1/2}$-$^2$P$_{3/2}$ & 2165.43$\pm$0.22& 0.07 & 0.22 &
 Feuchtgruber et al. 1997\\
    1997-Feb-20 & NGC\,6302 &    HI  & 7 - 5 & 2148.79$\pm \le$0.02& -0.06 & 0.14 &
 van Hoof \& Verner 1997\\
    1997-Feb-20 & NGC\,6302 &    [Na VII] & $^2$P$_{3/2}$-$^2$P$_{1/2}$ & 2134.60$\pm$0.15& 0.02 & 0.15 &
 Feuchtgruber et al. 1997\\
    1997-Feb-20 & NGC\,6302 &    HeII  & 8 - 7 & 2099.29$\pm \le$0.02& -0.07 & 0.13 &
 van Hoof \& Verner 1997\\
    1997-Feb-20 & NGC\,6302 &    [Mg V] & $^3$P$_1$-$^3$P$_2$ & 1782.58$\pm$0.20& 0.02& 0.20 &
 Feuchtgruber et al. 1997\\
    1997-Feb-20 & NGC\,6302 &    [Ar II] & $^2$P$_{1/2}$-$^2$P$_{3/2}$ & 1431.5831$\pm$0.0007& -0.11& 0.13&
 Yamada et al. 1985\\
    1997-Feb-20 & NGC\,6302 &    [S III] & $^3$P$_2$-$^3$P$_1$  & 534.387$\pm$0.005& 0.008 & 0.03 &
 Kelly \& Lacy 1995\\
    1997-Feb-20 & NGC\,6302 &    [Ar III] & $^3$P$_0$-$^3$P$_1$ & 458.081$\pm$0.006& 0.002 & 0.04 &
 Table 1\\
    1997-Feb-20 & NGC\,6302 &    [Ne V] & $^3$P$_2$-$^3$P$_1$ & 411.226$\pm$0.006& -0.007 & 0.04 &
 Feuchtgruber et al. 1997\\
    1997-Feb-20 & NGC\,6302 &    [O IV] & $^2$P$_{3/2}$-$^2$P$_{1/2}$ & 386.245$\pm$0.005& -0.001 & 0.03 &
 Feuchtgruber et al. 1997\\

    1996-Oct-19 & NGC\,7027 &    HI & 12 - 6 & 2284.95$\pm$0.02& -0.03 & 0.15 &
 van Hoof \& Verner 1997\\
    1996-Oct-19 & NGC\,7027 &    HI & 7 - 5 & 2148.79$\pm$0.02& 0.06 & 0.15 &
 van Hoof \& Verner 1997\\
    1996-Oct-19 & NGC\,7027 &    HI & 9 - 6 & 1692.56$\pm$0.02& 0.09 & 0.18 &
 van Hoof \& Verner 1997\\
    1996-Oct-19 & NGC\,7027 &    HI & 6 - 5 & 1340.51$\pm$0.01& -0.04 & 0.11 &
 van Hoof \& Verner 1997\\
    1996-Oct-19 & NGC\,7027 &    [Ar II] & $^2$P$_{1/2}$-$^2$P$_{3/2}$ & 1431.5831$\pm$0.0007& 0.00 & 0.13 &
 Yamada et al. 1985\\
    1996-Oct-19 & NGC\,7027 &    [S III] & $^3$P$_2$-$^3$P$_1$ & 534.387$\pm$0.005& 0.006 & 0.03 &
 Kelly \& Lacy 1995\\
    1996-Oct-19 & NGC\,7027 &    [Ne V] & $^3$P$_1$-$^3$P$_0$ & 411.226$\pm$0.006& -0.006 & 0.04 &
 Feuchtgruber et al. 1997\\
    1996-Oct-19 & NGC\,7027 &    [O IV] & $^2$P$_{3/2}$-$^2$P$_{1/2}$ & 386.245$\pm$0.005& 0.002 & 0.03 &
 Feuchtgruber et al. 1997\\
    1996-Oct-19 & NGC\,7027 &    [Ar V] & $^3$P$_{2}$-$^3$P$_{1}$ & 1265.57$\pm$0.04& -0.08 & 0.10 &
 Kelly \& Lacy 1995\\
    1996-Oct-19 & NGC\,7027 &    [Ar III] & $^3$P$_1$-$^3$P$_2$  & 1112.176$\pm$0.015& -0.01 & 0.07 &
 Kelly \& Lacy 1995\\
    1996-Oct-19 & NGC\,7027 &    [S IV]  & $^2$P$_{3/2}$-$^2$P$_{1/2}$  & 951.430$\pm$0.01& 0.013 & 0.05 &
 Serabyn, priv. com.\\

\tablenotetext{a}{Difference between observed and literature wavenumber of the 
                  reference line.}
\tablenotetext{b}{Measurement uncertainty in the SWS observation of the 
                  reference line.}

\enddata
\end{deluxetable}

\section{Conclusions}
Extending the results of paper I towards fainter lines, the ISO Short 
Wavelength Spectrometer has been used to obtain new rest wavelengths for 
7 mid-infrared fine structure lines. 6 of these are from species with lower 
ionization potentials $\geq40eV$, detectable in high excitation planetary 
nebulae and active galactic nuclei.

\acknowledgments
This work was suported by DARA grants 50 QI9402 3 and 50 QI 8610 8.


\end{document}